\documentclass[10pt,conference]{IEEEtran}
\IEEEoverridecommandlockouts
\usepackage{cite}
\usepackage{amsmath,amssymb,amsfonts}
\usepackage{graphicx}
\usepackage{textcomp}
\usepackage{hyperref}
\usepackage{float}
\usepackage{xcolor}
\usepackage[T1]{fontenc}
\usepackage{seqsplit}
\usepackage{caption}
\usepackage{makecell}
\usepackage{url}
\usepackage{listings}   
\usepackage{booktabs}
\def\BibTeX{{\rm B\kern-.05em{\sc i\kern-.025em b}\kern-.08em
    T\kern-.1667em\lower.7ex\hbox{E}\kern-.125emX}}
\begin{document}

\title{EvoChain: A Framework for Tracking and Visualizing Smart Contract Evolution}

\author{\IEEEauthorblockN{Ilham Qasse\IEEEauthorrefmark{1},
Mohammad Hamdaqa\IEEEauthorrefmark{1}\,\IEEEauthorrefmark{2}
Björn Þór Jónsson\IEEEauthorrefmark{1},
}

\IEEEauthorrefmark{1}Department of Computer Science,
Reykjavik University, Reykjavik, Iceland \\
\IEEEauthorrefmark{2}Department of Computer and Software Engineering, Polytechnique Montreal, Montreal, Canada \\
\IEEEauthorrefmark{1}\{ilham20$,$bjorn$,$mhamdaqa\}@ru.is, 
\IEEEauthorrefmark{2}mhamdaqa@polymtl.ca
}
\maketitle

\begin{abstract}
 Tracking the evolution of smart contracts is challenging due to their immutable nature and complex upgrade mechanisms. We introduce EvoChain, a comprehensive framework and dataset designed to track and visualize smart contract evolution. Building upon data from our previous empirical study, EvoChain models contract relationships using a Neo4j graph database and provides an interactive web interface for exploration. The framework consists of a data layer, an API layer, and a user interface layer. EvoChain allows stakeholders to analyze contract histories, upgrade paths, and associated vulnerabilities by leveraging these components. Our dataset encompasses approximately 1.3 million upgradeable proxies and nearly 15,000 historical versions, enhancing transparency and trust in blockchain ecosystems by providing an accessible platform for understanding smart contract evolution.
\end{abstract}


\section{Introduction}
Software evolution is a critical aspect of software development, enabling continuous improvement, adaptation to new requirements, and rectification of defects over time~\cite{mens2008introduction, Bennett2002}. Tracking software versions is essential for developers and stakeholders to understand changes, maintain compatibility, and ensure security throughout the software lifecycle~\cite{Spinellis2005, Conradi1998, Estublier2000}. In traditional software engineering, version control systems, and software repositories have long facilitated this process~\cite{Mockus2002}.

In blockchain technology and smart contracts, tracking software evolution introduces unique challenges. Smart contracts are self-executing code deployed on immutable blockchain platforms such as Ethereum~\cite{Szabo1996, buterin2014next, wang2019blockchain}. Once deployed, their code cannot be altered or upgraded in the conventional sense~\cite{hamdaqa2022icontractml,mohanta2018overview}. To enable updates, developers employ design patterns such as proxy contracts, which delegate calls to upgradeable logic contracts~\cite{salehi2022not, qasse2024immutable, ebrahimi2024large}. However, comprehending a smart contract's history, version dependencies, and associated vulnerabilities remains difficult due to the decentralized and transparent yet inherently complex nature of blockchain systems~\cite{Jiang2022, Atzei2017, Luu2016,soud2023automesc}.

The absence of comprehensive tools and datasets for tracking the evolution of smart contracts creates significant challenges. Developers and auditors lack efficient means to trace modifications across different contract versions, limiting their ability to monitor version changes. Additionally, assessing the impacts of upgrades implemented via proxy mechanisms on functionality and security is restricted. Finally, blockchain data's scattered and complex nature limits identifying and tracing the resolution of security vulnerabilities over time. These challenges impede efforts to ensure reliability, security, and transparency in smart contract ecosystems, ultimately affecting trust and adoption of blockchain technologies.

To address these challenges, we introduce EvoChain,\footnote{\url{https://github.com/IlhamQasse/EvoChain.git}} a novel framework designed to track and visualize the evolution of smart contracts. EvoChain integrates:
\begin{itemize}
    \item A comprehensive dataset, aggregating historical smart contract data, including code versions, deployment transactions, proxy relationships, and known vulnerabilities.
    \item Graph-based modeling, utilizing Neo4j to represent complex relationships between contracts, proxies, versions, and associated issues in a connected graph structure.
    \item An interactive visualization tool, offering a user-friendly web application that enables stakeholders to explore contract histories and dependencies without writing complex queries.
\end{itemize}

EvoChain offers significant contributions to the blockchain and software engineering communities. First, it enhances the understanding of smart contract evolution by systematically capturing and modeling contract versions and their relationships, aiding comprehension of their progression over time. Second, EvoChain facilitates security analysis by enabling the identification and tracing of vulnerabilities across contract versions, supporting efforts to improve contract security. Finally, it improves accessibility and transparency through an intuitive web interface that lowers the barrier for developers, auditors, and researchers to analyze smart contract evolution, promoting trust in decentralized applications.
By leveraging EvoChain, stakeholders can analyze smart contract upgrades, trace vulnerabilities, and facilitate greater transparency in blockchain ecosystems.

\section{EvoChain Overview}
EvoChain is a modular framework designed to track and visualize the evolution of smart contracts. EvoChain provides actionable insights into contract histories, upgrade motivations, and associated vulnerabilities by leveraging data from a prior empirical study, integrating graph-based modeling, and an interactive user interface. This section presents an overview of the architecture, encompassing the data layer, API layer, and user interface while detailing the methodology and its connection to prior work.
\subsection{Data Layer}
The data layer of EvoChain provides the foundation for tracking smart contract evolution. In this section, we discuss the data sources, the methodology used to collect and process the data, and the schema and storage mechanism employed to organize and query the dataset.
\subsubsection{Data Sources}
EvoChain builds upon a dataset derived from our previous empirical study~\cite{qasse2024immutable} which analyzed the lifecycle of upgradeable smart contracts on Ethereum. The two primary data sources for EvoChain are:
\begin{itemize}
    \item EthereumETL\footnote{\url{https://ethereum-etl.readthedocs.io/en/latest/}}: An open-source tool providing detailed historical data from the Ethereum blockchain, including blocks, transactions, logs, and events. This dataset serves as the foundation for analyzing interactions and upgrading patterns of smart contracts.
    \item Etherscan API\footnote{\url{https://etherscan.io}}: A widely used Ethereum block explorer providing verified smart contract source code, metadata, and additional details like creation timestamps and transaction counts. Etherscan complements the EthereumETL dataset by supplying off-chain information critical for code analysis and validation.
\end{itemize}
\subsubsection{Methodology Process}
The data collection process involves the following key steps:
\begin{enumerate}
    \item Filtering Upgradeable Contracts: In the study we employed PROXIFY,\footnote{\url{https://github.com/IlhamQasse/PROXiFY}} a tool designed to detect upgradeable smart contracts. PROXIFY analyzes contract bytecode and emitted events for patterns indicative of proxy usage, such as delegatecall operations and specific storage arrangements.
    \item Tracing Historical Versions:
    \begin{itemize}
        \item By examining emitted events (e.g., Upgraded, ImplementationUpdated) and transaction logs, EvoChain maps each proxy contract to the various implementation contracts it has managed over time, forming a complete upgrade history.
        \item Events containing addresses of new implementations are analyzed to reconstruct the sequence of versions associated with each proxy, revealing the relationships between proxies and their implementations.
    \end{itemize}
    \item Fetching Source Code and Metadata:
    \begin{itemize}
        \item Using the Etherscan API, EvoChain retrieves the verified source code for each contract version.
        \item  Additional metadata, including the contract's creation timestamp, transaction count, is gathered to contextualize each version.
    \end{itemize}
    \item Observed Changes in Smart Contracts: In our previous study~\cite{qasse2024immutable}, we categorized observed changes into four types: fixing vulnerabilities (addressing security issues), feature modifications (altering functionality through additions or deletions), gas optimizations (reducing gas costs for performance improvements), and other changes (minor adjustments or policy updates not fitting the other categories). We utilize tools such as Git diff\footnote{\url{https://git-scm.com/docs/git-diff}} to compare code differences between versions and the SmartBugs frameworkk\footnote{\url{https://hub.docker.com/u/smartbugs}} to detect security vulnerabilities. Gas cost comparisons are made using actual deployment data from Etherscan, providing a reliable overview of structural optimizations.
\end{enumerate}

\subsubsection{Data Schema}
\begin{figure}[t!]
    \centering
    \includegraphics[width=\columnwidth]{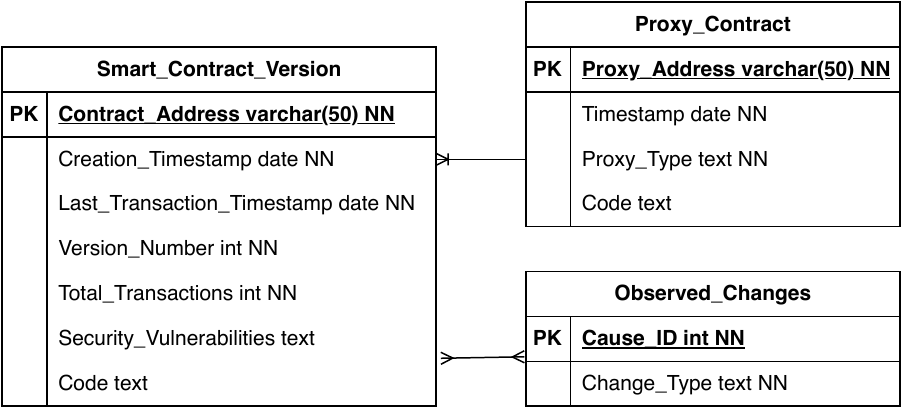} %
    \caption{EvoChain Data Schema}
    \label{fig:M}
\end{figure}

The dataset is structured as a graph in Neo4J, capturing the relationships between smart contract versions, proxies, and their upgrade paths. Core entities include:
\begin{itemize}
    \item Smart Contract Versions: Nodes representing contract versions, with attributes such as contract address, creation timestamp, last transaction timestamp, total transactions, version number, and vulnerabilities.
    \item Proxy Contracts: Nodes representing proxies, classified by type (e.g., EIP-1967, Transparent Proxy) and linked to the contracts they manage.
\end{itemize}
Key relationships include:
\begin{itemize}
    \item Implements: A one-to-many relationship linking a proxy contract to the smart contract versions it controls.
    \item Observed Changes: Annotates the nature of the change introduced in each version (e.g., security fix, performance optimization, feature addition).
\end{itemize}
Figure~\ref{fig:M} illustrates the graph schema, depicting the nodes, attributes, and relationships central to EvoChain.

\subsection{API Layer}
The API layer, built with Flask, serves as an intermediary between the data layer and the user interface, enabling seamless user interaction with the underlying data. Its key functions include:
\begin{itemize}
    \item Query Execution: Processes user inputs to retrieve data from the Neo4j database, such as contract relationships, version histories, and security information.
    \item Dynamic Data Retrieval: Integrates with the Etherscan API for real-time retrieval of verified source code and metadata, ensuring users have access to the most up-to-date information.
    \item Data Transformation: Structures raw query results into user-friendly formats suitable for visualization in the user interface.
\end{itemize}

This modular API design abstracts the complexity of interacting with graph databases and blockchain data, allowing users to focus on insights rather than implementation details.

\subsection{User Interface Layer}
The user interface provides an interactive platform for exploring the evolution of smart contracts, combining graphical and tabular views for comprehensive analysis. Key features include:
\begin{itemize}
    \item Graphical Visualization: EvoChain visualizes contract versions and proxy relationships as nodes and edges. Users can interact with the graph to trace upgrade paths, explore vulnerabilities, and identify root causes of changes.
    \item Tabular View: A synchronized tabular display lists detailed metadata for selected nodes, including creation timestamps, transaction counts, and vulnerabilities.
    \item Querying Capabilities: Users can search for contracts by address, proxy type, or version details, enabling flexible and targeted analysis.
    \item Real-Time Code Retrieval: The interface dynamically fetches verified source code from Etherscan, ensuring users can access the latest data.
\end{itemize}

The EvoChain dataset and tool are publicly available online\footnote{\url{https://github.com/IlhamQasse/EvoChain.git}} to support research and exploration of smart contract evolution.

\section{Applications of EvoChain}
EvoChain bridges a critical gap in the blockchain ecosystem by bringing the principles of version tracking and transparency, which are long established in traditional software development, to smart contracts. In conventional software engineering, version control systems like Git enable developers to track changes, collaborate effectively, and maintain a history of modifications [1]. EvoChain extends these capabilities to smart contracts, which traditionally lack such comprehensive evolution tracking due to the immutable nature of blockchain deployments.

One significant application of EvoChain is in enhancing transparency and user trust. By providing a clear and accessible history of smart contract versions, upgrades, and associated changes, users can make informed decisions about which contracts to interact with. This level of transparency is not typically available through standard blockchain explorers or tools, which often do not provide detailed insights into a contract's evolution or the reasons behind upgrades. EvoChain's visualization of upgrade paths and observed changes empowers users with knowledge about the contract's reliability and the responsiveness of developers to issues such as security vulnerabilities.

In security auditing and vulnerability analysis, EvoChain is a powerful tool for auditors and security professionals. By examining the relationships between contract versions and their associated vulnerabilities, analysts can track the remediation of known security issues over time. This continuous monitoring facilitates comprehensive risk assessments and supports the verification of adherence to security standards throughout a contract's lifecycle. EvoChain's ability to highlight unresolved vulnerabilities and the effectiveness of past upgrades enhances the efficiency of auditing processes and aids in proactive risk management.

For developers, EvoChain offers valuable insights into smart contract development and maintenance practices. Developers can learn from previous modifications by exploring upgrade paths and analyzing observed changes(such as bug fixes, feature additions, or gas optimizations). This knowledge helps in planning future upgrades more effectively, avoiding past mistakes, and adopting best practices. EvoChain's visualization of contract evolution aids in understanding the impact of changes, facilitating more strategic decision-making in the development process.

EvoChain also has significant applications in academic research and machine learning. Researchers can leverage the rich dataset provided by EvoChain to conduct large-scale empirical studies on upgrade patterns, security trends, and maintenance activities in smart contracts. This data can be instrumental in training machine learning models for various purposes, such as predicting potential vulnerabilities, assessing the risk of future upgrades, or forecasting contract evolution. By applying techniques like anomaly detection and pattern recognition, researchers can develop predictive analytics tools that enhance smart contracts' security and reliability.

Finally, in the context of investment decision-making, EvoChain enables investors and stakeholders to analyze the historical evolution of smart contracts. The transparency provided by the tool increases confidence in the reliability and integrity of contracts. Investors can estimate their potential investments' activity, longevity, and safety by assessing total transactions, contract age, and the history of security assessments. This informed approach aids in identifying trustworthy contracts and understanding the risks associated with interacting with specific smart contracts.
\section{Limitations and Future Work}
While EvoChain provides significant advancements in tracking smart contract evolution, certain limitations present opportunities for future enhancement. One of the primary limitations is the reliance on emitted events to detect upgrades. Since EvoChain depends on standardized upgrade events emitted by smart contracts, those that do not emit such events are not fully captured, leading to incomplete data. This reliance may overlook upgrades performed without event emissions or using unconventional methods, affecting the comprehensiveness of the dataset.

Another limitation is the focus on proxy-based upgrade patterns. While proxies are prevalent in enabling smart contract upgrades, they are not the only mechanism. EvoChain currently does not extensively analyze other upgrade approaches, such as data separation or strategy pattern. This narrow focus may introduce bias and limit insights into alternative upgrade practices within the blockchain ecosystem.
EvoChain's scope is also currently limited to the Ethereum blockchain, excluding contracts deployed on other platforms like Binance Smart Chain or Polkadot. This Ethereum-centric approach restricts the ability to compare practices and trends across different blockchain environments, potentially overlooking unique evolution patterns present in other ecosystems.

Scalability challenges present another limitation. Processing and storing the vast amount of data from the Ethereum blockchain poses difficulties in terms of data volume management and performance constraints. Maintaining responsive query performance and efficient data retrieval becomes increasingly complex as the dataset grows.

To address these limitations, future work on EvoChain will focus on several enhancements. One key improvement is the development of enhanced data collection methods. By incorporating alternative detection techniques, such as function call analysis or code similarity detection, EvoChain can identify upgrades beyond emitted events. This approach aims to capture a wider range of upgrade practices, improving the completeness of the dataset.

Another area of focus is promoting standardization within the developer community. By advocating for the adoption of standardized event naming conventions and logging practices, EvoChain can enhance the reliability of data collection and facilitate more accurate tracking of contract evolution. Collaboration with industry stakeholders and participation in standardization initiatives can drive this effort.

Expanding EvoChain's capabilities to include multiple blockchain platforms is another important direction for future work. Cross-platform support will enable comparative analyses and broaden the tool's applicability. Investigating how contracts evolve across different platforms will provide insights into best practices and common challenges, enriching the understanding of smart contract evolution globally.

Addressing scalability challenges involves implementing optimized data storage solutions and performance enhancements. Employing scalable architectures, database optimization techniques, and efficient caching strategies can manage large datasets effectively while maintaining responsive user interactions. These technical improvements will ensure that EvoChain remains a robust and user-friendly tool as it grows.

Lastly, including alternative upgrade mechanisms in the analysis will provide a more holistic view of smart contract evolution. By extending the scope beyond proxy patterns, EvoChain can analyze various upgrade methods, supporting more comprehensive insights and use cases.

By addressing these limitations and pursuing these enhancements, EvoChain aims to evolve into a more robust and versatile tool. These improvements will strengthen EvoChain's role in facilitating secure, transparent, and efficient smart contract development and maintenance, ultimately contributing to the advancement of blockchain technology.
\section{Related Work}
The detection and analysis of upgradeable proxy contracts have been the focus of several studies, employing methods such as source code analysis~\cite{bodell2023proxy,liu2024demystifying}, bytecode analysis ~\cite{huang2024sword,li2024characterizing}, and transaction history ~\cite{ebrahimiupc, salehi2022not}. While these studies advanced the understanding of proxy contracts, they primarily concentrated on detecting active proxies or their upgradeability features without focusing on comprehensively tracking historical versions.
Only a few studies have explored the evolution of smart contracts. Li et al.~\cite{li2024characterizing} and Liu et al.~\cite{liu2024demystifying} provided limited insights into historical versions, detecting 4,692 and 973 versions, respectively, but lacked a comprehensive view of changes over time, as summarized in Table~\ref{tab:related-work}.
In contrast, EvoChain surpasses these limitations by offering a comprehensive dataset and tool for tracking the evolution of smart contracts. It traces 14,990 historical versions across 1.3 million upgradeable proxies, significantly outscoring prior studies. Furthermore, EvoChain uniquely provides an interactive visualization tool, enabling users to explore smart contract evolution in depth. Unlike previous studies, EvoChain also makes its dataset publicly available, facilitating further research and enhancing transparency in the blockchain ecosystem.
\begin{table}[h!]
\centering
\caption{Summary of Studies on Upgradeable Proxy Contracts and Their Evolution}
\label{tab:related-work}
\begin{tabular}{lcc}
\toprule
\textbf{Study} & \textbf{\# Proxies Detected} & \textbf{\# Versions Tracked} \\ \midrule
\cite{salehi2022not}       & 8,225               & N/A                          \\
\cite{bodell2023proxy}     & 8,815               & N/A                          \\
\cite{ebrahimiupc}         & $\sim$3,000         & N/A                          \\
\cite{liu2024demystifying} & 44,282 (total)      & 973                          \\
\cite{li2024characterizing} & 43,650              & 4,692                        \\
\textbf{EvoChain}          & $\sim$1,300,000     & 14,990                       \\ \bottomrule
\end{tabular}
\end{table}
\section{Conclusion}
We presented EvoChain, a framework and dataset for tracking and visualizing smart contract evolution on Ethereum. By leveraging data from our previous study, modeling it in a Neo4j graph database, and providing an interactive web interface, EvoChain addresses challenges in understanding contract upgrades and vulnerabilities. The tool facilitates applications in security auditing, development insights, academic research, and investment decision-making. Despite limitations such as reliance on emitted events and focus on proxy patterns, EvoChain significantly enhances transparency and trust in blockchain ecosystems. Future work includes expanding to other blockchains, integrating predictive analytics, and improving data collection methods to capture a broader range of upgrade practices.
\bibliographystyle{ieeetr}
\bibliography{EvoChain}
\end{document}